\documentclass{article}
\usepackage{arxiv}
\usepackage{adjustbox}
\usepackage{multirow}
\usepackage[inline]{enumitem}
\usepackage{graphicx}
\usepackage{grffile}
\usepackage{subcaption}
\usepackage{tabularx}
\usepackage{xcolor}
\usepackage{hyperref}
\usepackage{comment}
\hypersetup{hidelinks}

\graphicspath{{images/}}
\newcommand{\Project}{MemShield}
\newcommand{\libProject}{libMemShield}
\newcommand{\libThread}{data thread}

\def\orcidID#1{ORCID iD: \texttt{https://orcid.org/{#1}}}
\def\authorrunning#1{\relax}
\def\titlerunning#1{\relax}
\def\email#1{\texttt{#1}}
\def\institute#1{\relax}
\def\inst#1{\relax}

\title{MemShield:\\GPU-assisted software memory encryption}
\date{December 16, 2019}

\begin{document}
\titlerunning{MemShield: GPU-assisted software memory encryption}
%
\newlength\authorsep\setlength{\authorsep}{.5em}
\author{%
\parbox{.47\textwidth}{\hfil
Pierpaolo Santucci\thanks{\orcidID{0000-0003-3596-7046}.}%
\hfil
Emiliano Ingrassia\thanks{\orcidID{0000-0002-6710-3392}.}%
\hfil} \\[\authorsep]
Epigenesys s.r.l., Rome, Italy\\[\authorsep]
\email{\{santucci,ingrassia\}@epigenesys.com}
\And
\parbox{.47\textwidth}{\hfil
Giulio Picierro\thanks{\orcidID{0000-0003-4079-8435}. Corresponding author.}%
\hfil
Marco Cesati\thanks{\orcidID{0000-0002-7492-3129}.}%
\hfil} \\[\authorsep]
University of Rome Tor Vergata, Rome, Italy \\[\authorsep]
\email{\{giulio.picierro,cesati\}@uniroma2.it} }
\maketitle              
\title{MemShield: GPU-assisted software memory encryption}
%
\begin{abstract}
Cryptographic algorithm implementations are vulnerable to Cold Boot attacks,
which
consist in exploiting the persistence of RAM cells across reboots or power down
cycles to read the memory contents and recover precious sensitive data. The
principal defensive weapon against Cold Boot attacks is memory encryption.  In
this work we propose  \Project, a memory encryption framework for user space
applications that exploits a GPU to safely store the master key and perform the
encryption/decryption operations.  We developed a prototype that is completely
transparent to existing applications and does not require changes to the
OS kernel. We discuss the design, the related works, the
implementation, the security analysis, and the performances of
\Project.
\end{abstract}


\let\oldfootnote\thefootnote
\renewcommand{\thefootnote}{\relax}
\footnote{\par\smallskip\leavevmode This article was accepted at the 18th
International Conference on Applied Cryptography and Network Security, ACNS
2020, October 19--22, 2020, Rome, Italy. The conference proceedings are
copyrighted by Springer and published in Lecture Notes in Computer Science
series. The final authenticated version is available online at
\texttt{https://doi.org/\emph{\small [to\,be\,assigned]}}.}
\let\thefootnote\oldfootnote

\section{Introduction}
Strong cryptographic algorithms rely on sensitive data that must be kept hidden
and confidential. Nowadays, the most practical attacks against data encryption
programs, full disk encryption layers (FDEs), authentication protocols, or
secure communication channels are based on memory disclosure techniques that
extract keys, hidden configurations, or unencrypted data directly from the RAM
memory cells of the running systems.  Consider, for an example, the (in)famous
OpenSSL's Heartbleed vulnerability \cite{heartbleed} that in 2014 allowed a
remote attacker to extract secret keys of X.509 certificates used for SSL/TLS
secure network protocols. 
Even worse, also properly implemented cryptographic programs are usually
vulnerable to the class of \emph{Cold Boot attacks} \cite{coldboothot}.
Basically, cryptographic programs rely on memory protection mechanisms
implemented by the operating system kernel to hide the sensitive data from any
unauthorized user or process in the system. The underlying assumptions,
however, are that (1) overriding the operating system control implies rebooting
the system, and (2) rebooting the system implies disrupting the contents of all
RAM cells.  While assumption (1) is correct if the operating system is
bug-free, unfortunately assumption (2) does not generally hold. Modern RAM
technologies, such as dynamic RAM (DRAM) and static RAM (SRAM), are based on
electric charges stored in small capacitors that could retain the charge for
several seconds, or even minutes, after power off. The typical Cold Boot
attack, therefore, consists of power cycling the system using the reset button
and quickly rebooting from a removable device into a program that extracts the
RAM contents \cite{coldboot}. Cold Boot attacks are so effective that they are
nowadays adopted even in digital forensic activities
\cite{carbone11,vomel2012}.

Many possible mitigations of Cold Boot attacks have been proposed in the past
few years, however there still does not exist a full, practical solution.
In fact, some proposals are based on cryptographic hardware circuits
integrated in the system, which are still not widely adopted
\cite{amdmet,inteltme,IntelSGX}. Other proposals are bound to specific hardware
architectures \cite{tresor,armored,henson2013beyond,softme}. Moreover, many
solutions are not completely transparent to the applications 
\cite{exzess,sugartaste,softme}. Finally, almost all proposed solutions
require changes to the operating system kernel
\cite{chen,cryptkeeper,ramcrypt,Huber,huber2019}.

In this work we introduce \emph{\Project}, a framework based on a
general-purpose Graphic Processing Unit (GPU) that encrypts the system memory
so as to mitigate the effects of Cold Boot attacks.  GPUs benefits of
widespread adoption and their massive parallelism may satisfy real-time demands
of transparent encryption systems, providing that communication overheads are
reasonable.  

\Project\ is designed so as to overcome most limitations of the previous
proposals. In particular, \Project\space
\begin{enumerate*}[label={(\roman*)}] 
    \item relies on a vanilla Linux kernel and does not require kernel patches;
    \item it is not bound to a specific hardware architecture;
    \item it does not require specific cryptographic hardware, and uses widely
        adopted GPUs as secure encryption key store and cryptographic
        processor;
    \item it can run on legacy systems;
    \item it stores encrypted data in system RAM, thus not limiting the amount
        of protected pages;
    \item it allows users to select the applications to be protected;
    \item it does not require source code changes in the protected applications;
    \item it exploits dynamic linking and does not require program
        recompilation;
    \item it achieves transparent memory accesses without code instrumentation;
    \item it uses a modular software architecture for the cryptographic
        components, thus permitting to easily change, enhance, or upgrade the
        encryption cipher;
    \item it is an open-source project, available at: {\small
        \url{https://gitlab.com/memshield/memshield/}}.
\end{enumerate*}

We developed a prototype based on the GNU/Linux operating system and CUDA GPUs.
It provides transparent memory encryption in user mode by using the userfaultfd
Linux kernel framework.  This prototype is a proof-of-concept aimed at testing
the validity of \Project's design, as well as assessing both security aspects
and performances.  As far as we know, \Project\ is the first framework that
achieves system memory encryption by using a GPU.  Moreover, while it relies on
a vanilla Linux kernel, it does not require patches to the kernel:
this is also a novelty for this kind of frameworks.

The article is organized as follows: in section~\ref{s:relwork} we compare
\Project\ with other solutions proposed in literature. In
sections~\ref{s:design} and~\ref{s:implem} we describe design and
implementation of \Project. In section~\ref{s:san} we discuss a security
analysis, while in section~\ref{s:perf} we illustrate
how \Project\ impairs the overall performances of the protected applications.
Finally, in section~\ref{s:conc}, we draw some conclusions.

\section{Related works}
\label{s:relwork}

Memory disclosure attacks, and in particular Cold Boot attacks
\cite{coldboot,bauer2016,coldboothot},  are nowadays a real menace that might
expose sensible information and precious data to unauthorized use or abuse.
Therefore, many researchers have proposed technical solutions to mitigate the
effects of this kind of attacks.

Some proposals are based on the assumption that the attacker might be
interested in a small portion of sensitive data, like the FDE master key used
for encrypting a physical disk partition, or the private key of an open SSL
connection. In these cases, an effective solution might be storing these data
outside RAM, that is, in volatile hardware registers. For example, an AES
encryption key could be stored in special CPU registers, such as x86\_64 SSE
registers (paranoix \cite{paranoix}, AESSE \cite{aesse}), debug registers
(TRESOR \cite{tresor}), profiling registers (Loop-Amnesia \cite{simmons11}), or
ARM NEON registers (ARMORED \cite{armored}).  Another approach is storing
sensitive data in locked-down static-RAM caches or hardware transactional
memories \cite{chaos2010,copker,guan2019}. However, storing sensitive data
outside system RAM does not scale up, so this solution is unfeasible when the
data to be protected have size larger than the storage provided by the
available volatile hardware registers.

An easy-to-implement mitigation for Cold Boot attacks aimed at large portions
of sensitive data might be based on the power-on firmware of the system. For
instance, during the initialization sequence the firmware might ask
for a user password, or it might erase the contents of all RAM cells
\cite{tcg}. However, these protections can be overridden, because it is
generally possible to change the firmware settings by resetting them to a
default configuration, for instance by inserting a jumper on the board or
removing the CMOS battery \cite{gruhn2016}. Even more crucially, the Cold Boot
attack might be executed by transplanting the memory chips in a suitable system
with a custom firmware \cite{vomel2011,vomel2012}. It has been shown
\cite{gruhn2016} that cooling down the memory chips to approximately 5~Celsius
degree before transplanting them allows an attacker to recover about 99\% of
the original bits. 

Thus, even if Cold Boot attacks can be mitigated, they cannot completely ruled
out: if resourceful attackers gains access to the physical RAM chips of a
running system, they are able to access the system memory contents.  Therefore,
the most effective protection against Cold Boot attacks is based on
\emph{memory encryption}, where the contents of some (or, possibly, all) memory
cells are encrypted by using one or more random keys stored in volatile
hardware registers outside of the RAM chips. In order to get meaningful
information from the memory dump, the attacker must either recover the
encryption key(s) from the volatile hardware registers, or break the encryption
cipher. 
Henson et al.~\cite{Henson} present a survey of hardware enhancements to
support memory encryption at memory controller, bus, and caches level,
protecting data transfers against eavesdropping.

Ideally, memory encryption could be implemented at hardware level so as to be
completely transparent to user mode applications and, in some cases, even to
the operating system kernel.  Nowadays, major chip manufacturers design
processors with MMU-based memory encryption. For instance, Intel Software Guard
Extensions (SGX) \cite{IntelSGX} allows user applications to create an
encrypted memory area, called \emph{enclave}, for both code and data. Encrypted
pages are stored in a page cache of limited size.
Confidentiality and integrity are guaranteed even when the OS kernel is
compromised.  However, SGX is not designed for full memory encryption and
legacy applications cannot use it without source code changes.  In 2017 Intel
announced Total Memory Encryption (TME)~\cite{inteltme}, a hardware extension
to encrypt all RAM cells with either a global key or multiple keys; in any
case, the encryption key cannot be retrieved by software.  TME is currently
under development.  AMD's proposed solution is named Secure Memory Encryption
(SME)~\cite{amdmet}: it encrypts and decrypts pages of memory if
a specific flag in the page table entries is set.  The encryption key is
randomly generated by the hardware and loaded at boot time into the memory
controller. An extension called TSME allows the hardware to perform
memory encryption transparently to the OS and the user software. Mofrad et
al.~\cite{sgxvssme} analyze features, security attacks, and performances of
both Intel SGX and AMD SME solutions.

While there is a raising interest to add support for MMU-based memory
encryption solutions in the operating system kernels \cite{linuxmeapi},
these systems are still not widespread, because this technology is
rather expensive and operating system support is scarce. Thus, in
the past years several researchers have proposed to encrypt the memory by
using frameworks exploiting either common-of-the-shelf (COTS) hardware
components or custom devices. Because the MMUs lack circuits aimed at
encryption, all these solutions must rely on procedures included in
hypervisors or operating system kernels, so they can be collectively named
\emph{software-assisted memory encryption} frameworks.
In this work we propose \Project, an innovative software solution based on COTS
hardware.

In software-assisted memory encryption, a typical access to a protected
page triggers a chain of events like the following:
\begin{enumerate*}[label={(\roman*)}]
\item the ordinary MMU circuits generate a hardware event (e.g., a
    missing page fault); \item the user application that made the access is
    suspended and an OS kernel/hypervisor handler schedules the execution of a
    cryptographic procedure (either on a CPU, or on some custom device like a
    FPGA or a GPU); \item the cryptographic procedure encrypts or decrypts  the
    data in the accessed page; \item the OS kernel/hypervisor resumes the
    execution of the user application.\end{enumerate*}
A crucial point of this mechanism is that the encryption keys used by
cryptographic ciphers must not be stored ``in clear'' on the system memory,
otherwise an attacker might retrieve them and decrypt all RAM
contents. Hence, software-assisted memory encryption proposals generally
include, as a component, one of the already mentioned solutions for storing a
limited amount of sensitive data outside of system RAM
\cite{tresor,simmons11,paranoix,chaos2010,copker,guan2019}.

We can reasonably expect that the performances of software-assisted memory
encryption are significantly worse than those of MMU-based memory encryption.
However, software-assisted memory encryption has potentially several advantages
over MMU-based memory encryption:
\begin{enumerate*}[label={(\roman*)}] 
    \item it can be used on legacy systems, as well as in low-level and
        mid-level modern systems without dedicated circuits in the MMU;
    \item it might be less expensive, because CPU/MMU circuits without
        cryptographic support are smaller and simpler;
    \item it is much easier to scrutiny its design and implementation in order
        to look for vulnerabilities and trapdoors;
    \item it is much easier to fix vulnerabilities and trapdoors, or to enhance
        the cryptographic procedures if the need arises.
\end{enumerate*}

One of the first proposal for software-assisted memory encryption, although
motivated by avoiding bus snooping rather than Cold Boot attacks, is in Chen et
al.\@ \cite{chen}: this framework uses locked-down static cache areas or
special scratchpad memories (usually found in embedded hardware) as a reserved
area for encrypted data.  This solution requires changes in the operating
system kernel to support memory access detection and to avoid that data in the
static cache are leaked into system RAM. The authors, however, do not discuss
how to protect the memory encryption key from disclosure.
A similar idea was explored in \cite{exzess}: the authors describe a FPGA
prototype named Exzess that implements a PCIe board acting as a transparent
memory proxy aimed at encryption.  The encrypted data are stored on the Exzess
device itself: device memory is mapped on the address space of a process, so
that read and write accesses on those pages trigger decryption and encryption
operations on the device, respectively. A drawback of this approach is that the
size of the ``encrypted RAM'' is limited in practice by the capacity of the
FPGA board.  \Project\ is quite different than these solutions  because it
stores the encrypted data on the system RAM.

CryptKeeper \cite{cryptkeeper} is a closed-source extension to the Linux kernel
swapping mechanism: user pages can be flagged as ``encrypted'' and, when their
number in RAM raises above a given threshold, removed from RAM and stored in
encrypted form in a swap area. However, CryptKeeper is fragile versus Cold Boot
attacks because it stores the encryption key in RAM.
A related idea is in Huber et al.~\cite{Huber}:
the authors suggest to perform encryption of user space processes memory at
suspend time, using the same key used for Full Disk Encryption (FDE).
Yet another variant of this idea is presented in \cite{huber2019}, where the
system memory of portable devices, like notebooks and smartphones, is encrypted
by means of the ``freezer'' infrastructure of the Linux kernel. \Project\ is
aimed at protecting against Cold Boot attacks possibly performed when the
system is up and running, so it has a rather different design.

Henson and Taylor \cite{henson2013beyond} describe a solution based on ARM
architecture that implements a microkernel exploiting a cryptographic
microprocessor to handle encrypted data either in a scratchpad memory or in the
system RAM. An improvement of this idea is in \cite{softme}, where the
ARM TrustZone ensures that unencrypted data never leak to system RAM. The
authors use the ARM Trusted Execution Environment to execute a
microkernel that loads encrypted program from RAM, decrypt them, and run
them safely.  The framework requires patches to the general-purpose OS kernel,
Linux, that handles the non-trusted programs.

RamCrypt~\cite{ramcrypt} is an open-source Linux kernel extension aimed at
encrypting memory at page-level granularity. It is transparent to the user
applications because the decryption is automatically triggered by the page
fault handler whenever an user access is attempted. RamCrypt protects anonymous
pages and non-shared mapped pages of the applications. In order to ensure
acceptable performances, a small number of the last recently accessed pages is
kept in clear in a ``sliding window'' data structure. RamCrypt also takes care
to avoid key leaks in RAM by using a slightly modified AES implementation from
TRESOR~\cite{tresor}.  Even if TRESOR has been shown to be vulnerable to some
classes of attacks \cite{tresorhunt}, RamCrypt is still an effective mitigation
against Cold Boot attacks. \Project\ adopts some ideas from RamCrypt, mainly
the encryption at page level triggered by page faults and the sliding window
mechanism; however, \Project\ does not require patches to operating system
kernel, and it makes use of a GPU as a safe store and processor for the
cryptographic operations.

HyperCrypt \cite{hypercrypt} is similar to RamCrypt, however memory encryption
is handled by a hypervisor (BitVisor) rather than a kernel
program; this allows HyperCrypt to also protect kernel pages.  Like RamCrypt,
HyperCrypt is based on TRESOR \cite{tresor}. TransCrypt \cite{transcrypt} is
similar to HyperCrypt, yet it relies on ARM Virtualization Extensions to
implement a tiny encryption hypervisor.

Papadopoulos et al. \cite{sugartaste} proposed a framework for
software-assisted memory encryption that can protect either part of, or the
whole system memory; the master key is kept in CPU debug registers, like in
\cite{tresor}.  The framework relies on code instrumentation to intercept
load/store instructions that access memory, and it requires some patches to the
operating system kernel.

Finally, EncExec \cite{encexec} makes use of static caches as storage
units for encryption keys and unencrypted data. Whenever a user application
accesses an encrypted page, EncExec decrypts the data and locks them in the
static cache so that the application can transparently get the unencrypted
data. EncExec adopts an interesting approach, however it requires patches to
the operating system kernel.

Concerning the usage of the GPU for safely executing cryptographic routines,
the seminal work is PixelVault~\cite{pixelvault}, which implements AES and
RSA algorithms on the GPU by carefully storing the cryptographic keys 
inside the GPU registers. The main goal of that work is to implement a more
secure version of the OpenSSL library in order to mitigate memory disclosure
attacks.
PixelVault is based on a GPU kernel that runs forever waiting for crypto
operation requests submitted via a memory regions shared between CPU and GPU.
As discussed in \cite{Zhu}, PixelVault is actually vulnerable to unprivileged
attackers, however several authors suggested ways to enhance its approach. For
instance, the authors in \cite{wang2018} propose to run CUDA cryptographic
applications inside guest VMs by using a virtualized GPU; no privileged
attacker on guest VMs is able to retrieve the encryption keys, because they are
never stored in the guest VM memory. In \cite{graviton} the authors suggest to
modify the GPU hardware to prevent the device driver from directly accessing
GPU critical internal resources.  In \cite{jang2019} the authors propose to
use a custom interface between GPU and CPU and to extend the Intel SGX
technology to execute the GPU device driver in a trusted environment that a
privileged attacker cannot access.
\Project{} mitigates Cold Boot attacks, so
it assumes that privileged users are trusted (consider that a privileged user
might easily get a full memory dump without a Cold Boot attack).  Therefore,
\Project\ just ensures that unprivileged users cannot access the GPU, thus
avoiding the original PixelVault vulnerabilities.

\section{Design overview}
\label{s:design}

\Project\ is designed to mitigate Cold Boot attacks perpetrated by malicious
actors with physical access to the machine. We assume trusted privileged users
and processes, as well as safe operating system and base programs.

The rationale behind \Project{}'s design was to provide memory encryption
services to multiple concurrent users logged on the same machine, using a COTS
GPU as secure key store and cryptographic processor. Thus, \Project\ does not
rely on custom cryptographic circuits or specific hardware architectures.

GPUs are massively parallel accelerators consisting in thousand of cores
typically grouped in several \emph{compute units}. Cores in the same compute
unit can communicate via \emph{shared memory}, typically implemented as a fast
user programmable cache memory. Different compute units can communicate each
other via \emph{global memory}, that is, the GPU's RAM. In GPU programming
terminology, CPU and system RAM are referred as \emph{host}, while GPU and its
RAM are referred as \emph{device}. Device code is written in special functions
called \emph{kernels}. Those functions, once invoked from the host, can trigger
multiple parallel executions of the same kernel function over the input,
depending on how the kernel is launched.

\Project{} transparently encrypts user-space memory at page granularity and
decrypts them on-demand whenever an access is attempted. There is no need
to change or rebuild the source code of the protected applications.
Moreover, users may select the programs to be protected, so that applications
that do not handle sensitive data do not suffer any slowdown.

By design, for each process and at any time, \Project{} enforces a bounded
number of pages in clear, while keeping most of them encrypted. This mechanism
is based on a \emph{sliding window} and it is an effective solution against
Cold Boot attacks, as already proved in \cite{ramcrypt}. Encrypted data are
stored in system RAM, hence there is virtually no limit on the amount of pages
that can be protected.

\Project{}'s core is a daemon that is in charge of encrypting and decrypting
memory on behalf of clients.  We define as \emph{client} any process interested
in memory encryption services.  Because GPU programming is supported by user
mode tools and libraries, \Project{} daemon runs in user mode.

To support transparent memory encryption, the client that attempts to access an
encrypted page must be suspended and decrypted contents for that page must be
provided.  In order to achieve transparent memory encryption,
\Project{} must be able to detect clients' attempts to access encrypted memory
and provide decrypted contents.  As a matter of fact, detection of memory
accesses represents one of the most challenging aspects of the project.
We chose to address this issue by using userfaultfd \cite{KERNuserfaultfd},
a framework recently added to the Linux kernel aimed at efficient page fault
handling in user space. Thanks to userfaultfd, no changes are required to
the operating system kernel.
Currently userfaultfd is used by applications that implement memory
checkpoint/restore functionality~\cite{criu} and live migration of both
containers~\cite{stoyanov} and virtual machines~\cite{suetake}. \Project\  is
the first project that uses userfaultfd for memory encryption, as far as we
know.

In systems with a Memory Management Unit (MMU), the \emph{logical addresses}
appearing in CPU instructions must be translated in \emph{physical addresses}
to be used for RAM's chips programming. Translations are described by means of
the \emph{page tables}, usually a per-process data structure kept by the OS
kernel.
Each page table entry contains a page's virtual-to-physical mapping and
attributes such as protection bits (Read, Write, eXecute).  A \emph{page fault}
is generated by the MMU hardware if a translation is not found (\emph{missing
fault}) or the wrong access type is attempted (\emph{protection fault}).
This mechanism is actually used by operating systems to isolate process address
spaces.
In all POSIX-compliant systems there exists an established technique to detect
memory accesses in user space: it consists of changing the page permissions so
as to forbid any access, then executing a custom handler for the SIGSEGV
signal sent by the kernel to the process at any access attempt
\cite{edelson1992,cesati2015}.  However, this mechanism may significantly
impair the performances of the clients compared to userfaultfd.

A crucial design goal of \Project\ is transparency with respect to its clients:
no change to the source code of the clients is required, and no recompilation
is needed.  In order to achieve this goal, we assume that clients are
dynamically linked to the C library (\emph{libc}); in practice, this is almost
always true.
The idea is to intercept calls to the memory management
functions typically provided by the \emph{libc} implementation, in order to
register the client's memory areas to userfaultfd.  The custom handlers of the
memory management functions are collected in a user-mode library called
\mbox{\emph{\libProject}}. This library can be loaded in the client's process
before the C library using the \texttt{LD\_PRELOAD} technique, so
that the custom handlers override the original library symbols.

Because \Project\  must handle several concurrent clients with a single GPU,
userfaultfd works in non-cooperative mode: a single user-mode process is
notified of any page fault events occurring in registered clients. We call this
process the \Project's \emph{server}.

The server does not perform cryptographic operations directly, because it
cannot access the encryption keys stored in the GPU registers. The most
straightforward way to implement this mechanism is to launch on the GPU an
always-running kernel that implements a safe cipher.

\subsection{Design limitations}

By design, \Project\ handles only private anonymous memory: it is not concerned
with memory areas backed by files or shared among different processes. It is
also based on userfaultfd, which has some constraints of its own: mainly, it
cannot protect the memory area handled by \texttt{brk()} or \texttt{sbrk()}.
However, \Project\ can handle the client's stack, any memory area obtained by
\texttt{malloc()}, \texttt{calloc()}, \texttt{realloc()}, as well as the
anonymous memory obtained by \texttt{mmap()} and \texttt{remap()}: typically,
sensitive data end up being stored in such pages.

%

\section{Implementation details}
\label{s:implem}

The main activities performed by \Project\ are:
\begin{enumerate*}[label={(\roman*)}]
    \item memory area registration to userfaultfd,
    \item page fault handling,
    \item sliding window management, and
    \item GPU cryptographic operations.
\end{enumerate*}

\subsection{Memory area registration}

When a client allocates a memory area to be protected, \Project\ must register
the corresponding set of virtual addresses to userfaultfd.  In order to
achieve this, \libProject\ overrides some C library functions.

For allocations performed by anonymous memory mapping (\texttt{mmap()} and
analog functions), the custom wrapper just performs the original procedure
and registers the obtained virtual address to userfaultfd.

On the other hand, handling memory areas obtained by \texttt{malloc()} and
similar functions is more demanding, because the C library might use the
\texttt{brk()}/\texttt{sbrk()} system calls to perform the allocation.
\Project\ forces the C library to always use anonymous memory mapping when
allocating memory by means of the \texttt{mallopt()} tuning
function.

The userfaultfd framework does not handle stack pages. To overcome this
problem, \libProject{} replaces original stack pages with memory allocated
with \texttt{mmap()} using \texttt{sigaltstack()}. Since stack encryption
could have a significant impact on overall performances of \Project{},
encrypting stack pages can be selectively enabled or disabled by means of
an environment variable.

\libProject\ also overrides the \texttt{free()} and \texttt{munmap()}
functions, because it ensures that any page is filled with zeros before
releasing it to the system. Of course, this is crucial to avoid leaking
sensitive data in RAM.

\subsection{Page fault handling}

At the core of \Project\ there is the virtual address space mechanism
implemented by the operating system.  Any client is allowed to access a given
number of anonymous pages ``in clear'', that is, in unencrypted form.  These
pages belong to the address space of the client and the corresponding physical
page frames are referenced in the page tables. On the other hand, the page
table entries associated with encrypted pages denote missing physical pages.
The physical pages storing the encrypted data belong to the server's address
space, and are referenced in a per-client red-black tree sorted by virtual
addresses.

When a client accesses an encrypted page, the MMU raises a page fault because
of the missing physical page in the client's page table. Since the
corresponding virtual address has been registered to userfaultfd, the server is
notified about the event. As a consequence, the server looks for the virtual
address of the missing page in the client's red-black tree and retrieves the
physical page containing the encrypted data.  Then, the server sends the
encrypted data to the GPU, which performs the decryption operation.
Subsequently, the server relies on userfaultfd to resolve the client's page
fault by providing a physical page containing the decrypted data.

If the server does not find a virtual address in a client's red-black tree, the
page is missing because it has never been accessed earlier. Therefore, the
server allocates a new red-black tree node and resolves the client's page fault
by providing a physical page containing all zeros.

\subsection{Sliding window management}

The server keeps a per-client data structure called sliding window, which is a
list of virtual addresses corresponding to unencrypted anonymous pages of the
client. The sliding window maximum size is configurable.
When the server is going to resolve a page fault, it adds the corresponding
virtual address to the sliding window.  It also checks whether its size has
become greater than a preconfigured maximum. In this case, it takes away the
oldest unencrypted page in the sliding window, which is thus removed from the
client's page tables.

The server cannot operate on the address space of the client. Therefore,
\libProject\ creates at initialization time a thread called \emph{\libThread},
which acts on the client's address space on behalf of the server.
Correspondingly, a thread is created in the server to handle
the requests of the client concurrently with those of the other clients.

Server and \libThread\ exchange information by means of a Unix socket and a
shared memory area.  The Unix socket is used only to transmit open file
descriptors and control messages.  \Project\ does not send sensitive data with
this socket, because the Linux implementation makes use of a kernel memory
buffer which could be vulnerable to Cold Boot attacks. The shared memory area,
instead, is safe, because it is composed by a single page frame in RAM, which
is explicitly filled with zeros by the server as soon as a transfer is
completed.

When the server must drop a page from a sliding window,
it sends the corresponding virtual address to the \libThread. The latter
provides the contents of the page, that is, the unencrypted data, to the
server; then, it clears the physical pages and invokes the
\texttt{MADV\_DONTNEED} command of the \texttt{madvise()} system call to remove
the physical page from the page tables. The server encrypts the data by means
of the GPU, and adds the encrypted page to the client's red-black
tree.

\subsection{GPU encryption}

\Project{} cryptographic module services the requests coming from the server
using the GPU both as a secure key store and a secure processor. Any
cryptographic procedure operates on a single page of 4096 bytes.

The data are encrypted by using the cipher ChaCha~\cite{chacha}. This choice
was motivated by the need for a cipher that is, at the same time,
\begin{enumerate*}[label={(\roman*)}] \item secure, \item suitable for GPU
 computation, and \item simple enough so that the computation can be
 performed completely in GPU registers.\end{enumerate*} \Project{} implements
the strongest variant ChaCha20, which can be regarded as cryptographically
safe \cite{maitra2016chosen,dey2017improved,choudhuri2016differential}.

The actual encryption/decryption of a 4096-byte page is performed by
\texttt{XOR}ing the data with several keystream blocks.  ChaCha20 computes a
512-bit block of keystream starting from a 384-bit seed composed by a 256-bit
key and a 128-bit value. In \Project\ the 256-bit key is unique and it is
generated by a cryptographically secure pseudo-random number generator provided
by the operating system. This key is sent to the GPU and stored only in GPU
registers, afterwards it is purged out of the server memory.

The 128-bit value of the seed, which can be used both as a counter in stream
ciphering and as a predefined nonce, is composed by the virtual address of the
page, the process identifier (PID) of the client, and a counter ranging from 0
to 63 that corresponds to the index of the 512-bit block inside the page.
Observe that the keystream blocks could be generated and \texttt{XOR}ed
independently with the plaintext blocks. The ciphertext construction is thus
embarrassingly parallel, which is a highly desirable feature in a GPU
implementation. Another useful property is that encryption and decryption are
performed with the same operations, hence the GPU kernel can use the same
function for both.

\Project{} cryptographic module makes use of a NVIDIA GPU programmed by means
of the CUDA toolchain \cite{CUDA2008,cudatk}. The GPU is reserved to \Project,
which means that unprivileged users cannot access the device.  In practice,
because the communication channel between user space and the CUDA driver is
based on device files, the permissions of these device files are changed so
that access is only allowed to privileged users.


The GPU kernel consists of several CUDA blocks; each block acts as a
\emph{worker} whose job is to extract pages from a queue and process them using
32 CUDA threads (one \emph{warp}). Each CUDA thread generates two ChaCha20
keystream blocks (128 bytes), which are then \texttt{XOR}ed with the same
amount of plaintext. The number of CUDA blocks in the GPU kernel is dynamically
computed at run time according to the features of the GPU board. Using more
than one block allows the server to submit requests for several concurrent
clients.

Because the 256-bit encryption key is created at initialization time and stored
inside the GPU registers, the GPU kernel cannot be terminated, otherwise the
key would be lost. \Project\ uses a mapped memory between host and device to
implement a shared data structure that controls the GPU operations.
The data transfer between host (server) and device (GPU) is realized through a
circular buffer implementing a multiple-producer, single-consumer queue:
multiple host threads can submit concurrent pages to the same queue, while
those will be processed by a single worker on the GPU. Each worker has its own
queue, thus the workers runs independently and concurrently.

An important aspect of the implementation is that the encryption key and the
internal state of the cipher are never stored in GPU local memory, otherwise
\Project\ would be vulnerable to GPU memory disclosure attacks. We verified
that the current implementation of the GPU kernel never does register spilling.

\subsection{Prototype limitations}

The current implementation of \Project{} is a prototype, thus it has some
limitations. First of all, any protected application must be single process.
There is no major obstacle to enhance \Project\ so as to overcome this limit.

Protected applications must also be single thread. It would be possible to
extend \Project\ to support multi-threaded processes whenever userfaultfd
becomes capable of handling write-protect page faults. Work is in progress
to integrate this feature in the vanilla Linux kernel \cite{aa.git}.

\Project\ protects all private anonymous pages. In order to have better
performances, it could be preferable to selectively encrypt only the subset of
pages containing sensitive data.

Finally, the ChaCha20 implementation is prototypal and could 
be improved.

\section{Security Analysis}
\label{s:san}

\Project\ is an effective mitigation against Cold Boot attacks. In fact,
RamCrypt's authors already proved \cite{ramcrypt} that the sliding window
mechanism is an effective technique that could drastically reduce the
probability to find meaningful encryption keys or other sensitive data in
memory dumps. Like RamCrypt, \Project\ makes use of one sliding window per each
protected application.

\Project\ is also inspired by how PixelVault \cite{pixelvault} makes use of the
GPU to safely store encryption keys and run cryptographic procedures. However,
PixelVault is nowadays assumed to be vulnerable \cite{Zhu,cudaleaks}. The
reported attack vectors to PixelVault were based on launching malicious kernel
functions on the GPU, or running a CUDA debugger on the running GPU kernels.
\Project\ avoids these vulnerabilities because it restricts access to the GPU
to privileged users. Recall that, in our threat model, privileged users and
privileged processes are always regarded as trusted. \Project\ also takes care
of avoiding GPU register spilling, so that Cold Boot attacks against the GPU
memory would not retrieve any sensitive data of the cryptographic procedure.

Memory dumps obtained by Cold Boot attacks might expose data included in the
kernel buffers associated to Unix sockets, pipes, or other process
communication channels. Actually, we verified that the Linux implementation of
Unix sockets is vulnerable, because the kernel never erases the associated
buffers. \Project\ carefully avoids sending sensitive data by means of
Unix sockets. Rather, it makes use of shared memory, whose contents can be
explicitly cleared by \Project\ at the end of the sensitive data transmission.

\Project\ does not weaken the operating system
isolation guarantees at runtime, thus the confidentiality of users' data
against malicious users logged on the same system is preserved.
In particular, observe that a unprivileged user cannot tamper with the server
daemon, because we assumed that the operating system is safe and the server
is a privileged process.

A malicious user could try to interfere with \Project's communication channels.
The daemon listens on a Unix socket waiting for connection requests from
clients; hence, unprivileged processes must have write permissions on this
socket.  However, an attacker cannot replace the socket (in order to mount a
man-in-the-middle attack), because the socket interface file is placed in a
directory owned by root and not writable by unprivileged users.

A malicious user cannot even tamper with the shared memory area used to
exchange data between client and daemon: in fact, this area is created by means
of file descriptors exchanged privately through the Unix socket; the area is
then mapped in the address spaces of both daemon and client.
%

\Project\ is not designed to protect sensitive data against DMA attacks or
other side channel attacks; just consider that a successful DMA attack might
break the whole operating system guarantees, while in our threat model the
operating system is assumed to be sound.

\section{Performance Evaluation}
\label{s:perf}

In order to establish the performance impact of \Project\ prototype on
protected applications, we ran some benchmarks.  All tests have been executed
on a workstation equipped with a 3.5~GHz Intel Core i7 4771 CPU having 4
physical and 8 logical cores, 32~GiB RAM, and a GPU NVIDIA GeForce GTX 970
(compute capability 5.2) with 4~GiB of device memory.  The workstation used
Slackware 14.2 as Linux distribution with kernel 4.14, \emph{glibc} version
2.23, NVIDIA driver version 418.67 and CUDA 10.1.  CPU power-saving was
disabled.

\begin{table}[b]
	\centering
	\begin{tabular}{|cc|r|r|r|r|}
\cline{3-6}
\multicolumn{2}{c|}{}  & \multicolumn{1}{c}{\textbf{aes}} & \multicolumn{1}{c}{\textbf{sha512}} & \multicolumn{1}{c}{\textbf{qsort}} & \multicolumn{1}{c|}{\textbf{crypt}}\\
    \hline
\multicolumn{2}{|c|}{\textbf{Baseline}} & 0.30 \(\pm\) 0.03 & 0.44 \(\pm\) 0.01 & 4.67 \(\pm\) 0.03 & 4.76 \(\pm\) 0.02\\
\cline{1-2}
 \multirow{4}{*}{~\rotatebox{90}{\parbox{4em}{\textbf{\,Sliding window}}}} & \textbf{32} & 1.76 \(\pm\) 0.02 & 0.47 \(\pm\) 0.01 & 6.07 \(\pm\) 0.02 & 12.53 \(\pm\) 0.08\\
    &\multirow{1}{*}{\textbf{16}} & 1.94 \(\pm\) 0.01 & 0.48 \(\pm\) 0.00 &  6.08 \(\pm\) 0.04 & 12.56 \(\pm\) 0.09 \\
    &\multirow{1}{*}{\textbf{8}} & 16.64 \(\pm\) 0.09 & 4.39 \(\pm\) 0.02 &  6.10 \(\pm\) 0.03 & 12.56 \(\pm\) 0.10 \\
    &\multirow{1}{*}{\textbf{4}} & 35.23 \(\pm\) 0.22 & 4.58 \(\pm\) 0.03 & 64.35 \(\pm\) 1.62 & 12.52 \(\pm\) 0.08\\
    \hline
    \end{tabular}

    \vskip1em
	\caption{%
    Wall-clock execution times of single-instance benchmarks, with different
    sliding window sizes and without \Project{} protection (``Baseline'').
    Average times and standard deviations are in seconds.  Sliding window sizes
    are in pages. \Project\ encrypts all private anonymous pages, including the
    stack.
	}
	\label{t:slw}
\end{table}

Two benchmarks (\emph{crypt}, and \emph{qsort}) belong to
the stress-ng~\cite{stressng} test suite, version 0.09.48.
\emph{crypt} consists of multiple invocations of the C library
\texttt{crypt\_r()} function on data placed onto the stack. The test is
executed by:
{\footnotesize\begin{verbatim}
$ stress-ng --crypt 1 --crypt-ops 1000
\end{verbatim}}
\noindent
\emph{qsort} performs multiple sorts of an array of elements allocated with
\texttt{calloc()}. To sort the array, the C library function \texttt{qsort()} is
used:
{\footnotesize\begin{verbatim}
$ stress-ng --qsort 1 --qsort-size 2048 --qsort-opts 10000
\end{verbatim}}
\noindent
The third benchmark is \emph{aes} from \emph{OpenSSL} suite version~1.0.2s,
which operates on data structures stored in pages allocated with several calls
to \texttt{malloc()} interleaved with calls to \texttt{free()}. The test
consists of encrypting the Linux kernel~5.0 source archive, placed in a
RAM-based filesystem, using AES-256 in CBC mode:
{\footnotesize\begin{verbatim}
$ openssl enc -aes-256-cbc -in linux-5.0.tar.gz -out /dev/null -k pass
\end{verbatim}}
\noindent
Finally, the fourth benchmark is \emph{sha512} from \emph{GNU Coreutils}~1.25:
it invokes \texttt{malloc()} to allocate a single buffer storing file data,
then it computes the digest by storing cryptographic internal state on the
stack. The test was launched on the Linux kernel~5.0 source archive placed in a
RAM-based filesystem:
{\footnotesize\begin{verbatim}
$ sha512sum linux-5.0.tar.gz
\end{verbatim}}

In every test run we collected the execution times by using GNU
\emph{time}~1.7; each specific test has been repeated 10 times, then average
values and standard deviations have been computed.

\begin{figure}[t]
	\centering
	\begin{subfigure}[b]{0.49\textwidth}
		\includegraphics[width=\textwidth]{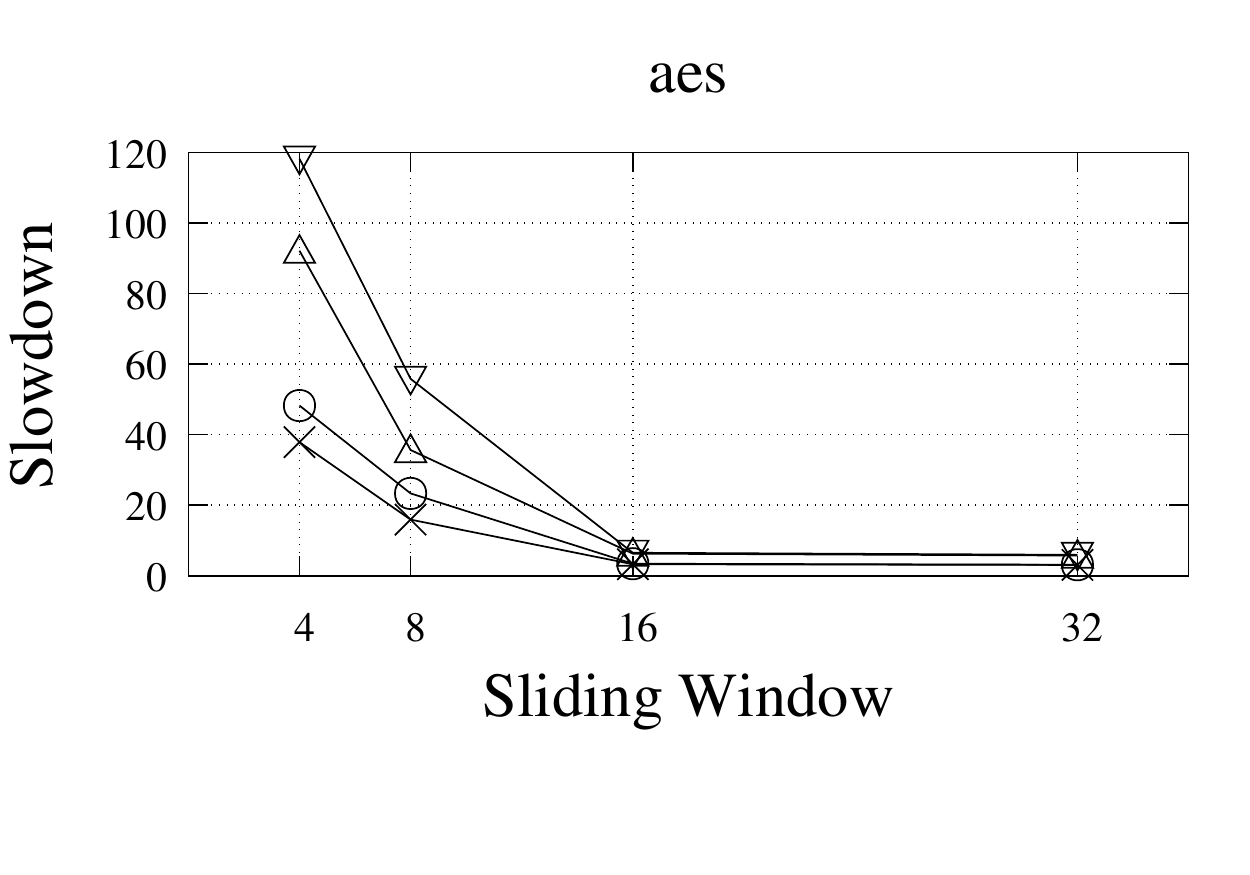}
		\label{fig:aes:sw}
	\end{subfigure}
	\begin{subfigure}[b]{0.49\textwidth}
		\includegraphics[width=\textwidth]{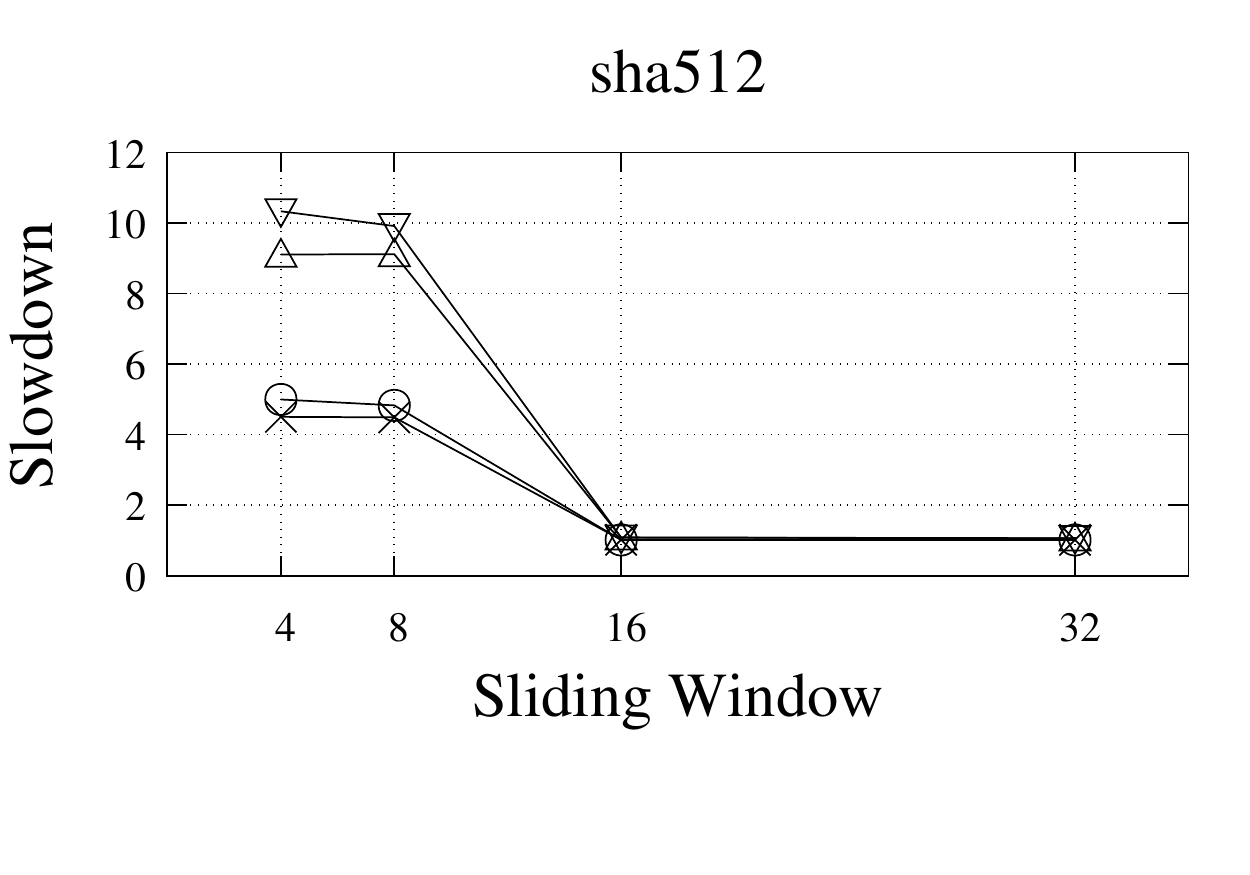}
		\label{fig:sha512:sw}
	\end{subfigure}
    \vskip-2.5em\includegraphics[height=1.9em]{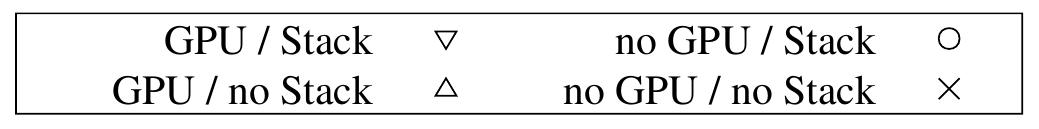}\vskip-.5em
	\begin{subfigure}[b]{0.49\textwidth}
		\includegraphics[width=\textwidth]{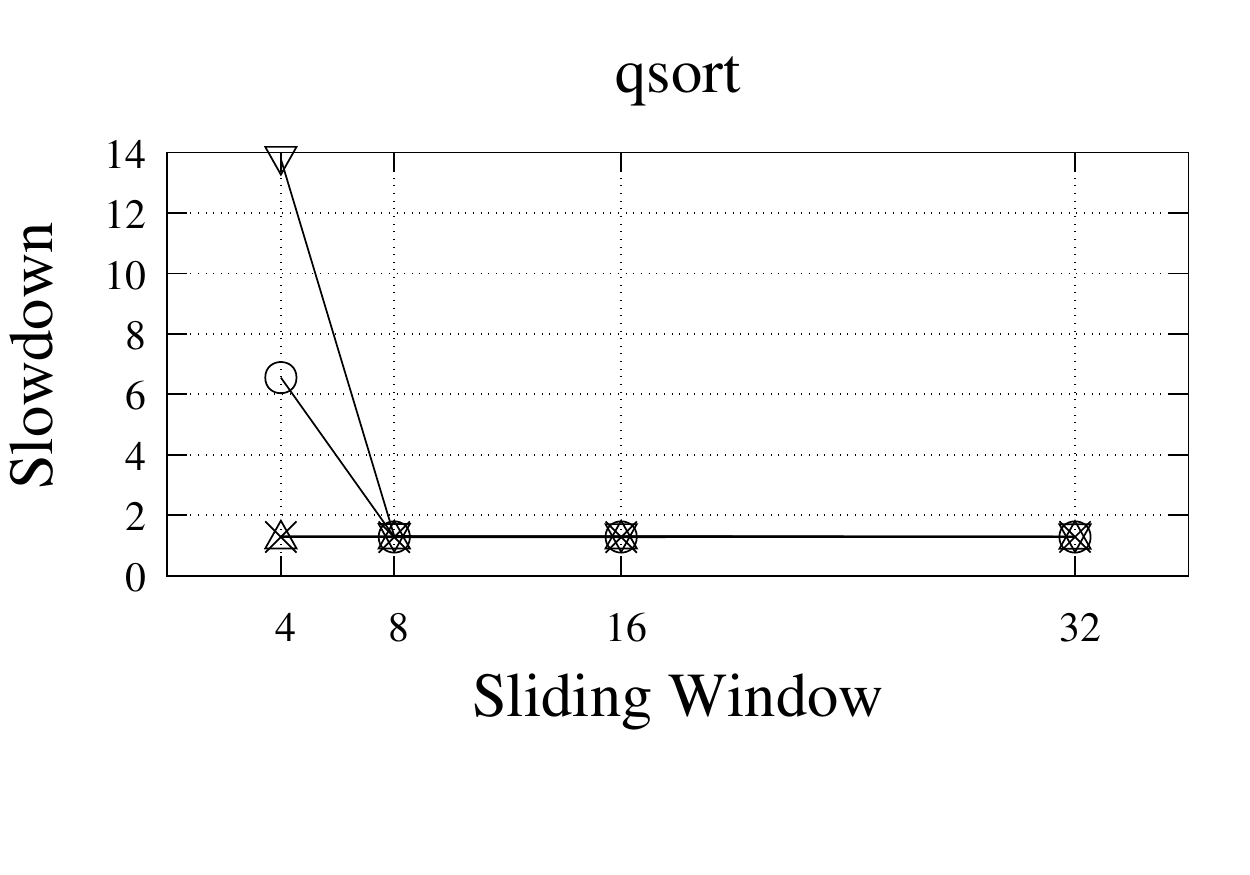}
		\label{fig:qsort:sw}
	\end{subfigure}
	\begin{subfigure}[b]{0.49\textwidth}
		\includegraphics[width=\textwidth]{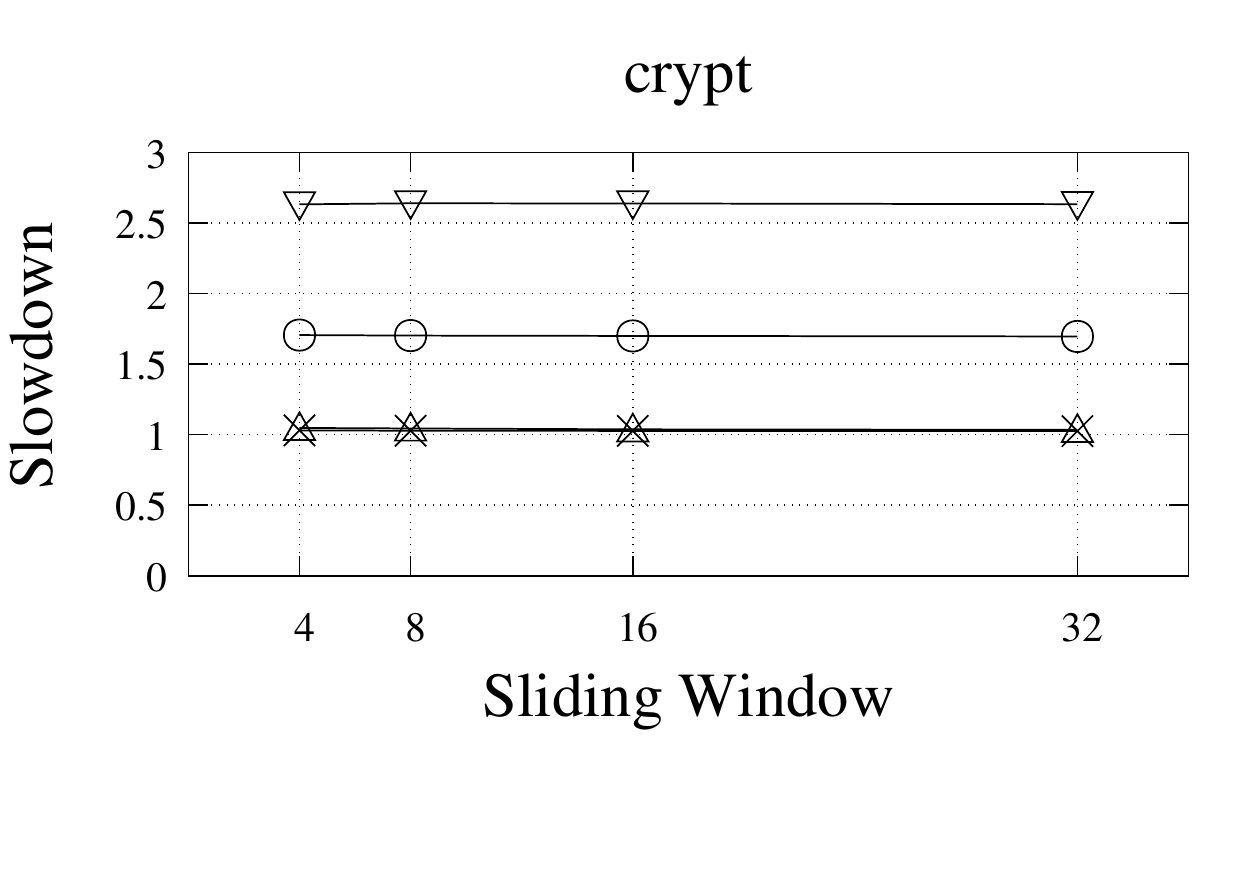}
		\label{fig:crypt:sw}
	\end{subfigure}
    \vskip-3em
    \caption{Slowdowns of a single instance of the benchmarks relative to
    baseline (no \Project) varying the sliding window size.  \Project\ protects
    all private anonymous pages including the stack (``Stack'') or excluding it
    (``no Stack'').  Encryption is performed by the GPU (``GPU'') or not
    done at all (``no GPU'').}
	\label{fig:sw}
\end{figure}

Table~\ref{t:slw} reports how \Project\ affects the average execution times of
the four benchmarks with different sliding window configurations.
Figure~\ref{fig:sw} shows the slowdowns of the four benchmarks with respect to
the baseline, which is the running time without \Project\ protection. In order
to better understand how the different components of \Project\ contribute to
the overhead, any plot has four lines, which correspond to the following cases: 
\begin{enumerate*}[label={(\roman*)}]
    \item encryption on GPU of all private anonymous pages, including the stack
        (``GPU, Stack''),
    \item encryption on GPU of all private anonymous pages, excluding the stack
        (``GPU, no Stack''),
    \item no encryption at all (the GPU is not involved in handling the
        protected pages, thus \Project\ server stores the pages in clear), but
        handling of all private anonymous pages, including the stack (``no GPU,
        Stack''), and finally
    \item no encryption at all, for all private anonymous pages, excluding the
        stack (``no GPU, no Stack'').
\end{enumerate*}
Distinguishing between ``Stack'' and ``no
Stack'' slowdowns is important because, when an application does not have
sensitive data stored on the stack, disabling stack encryption significantly
improves the performances of \Project. Distinguishing between ``GPU'' and ``no
GPU'' slowdowns is useful in order to understand how much the userfaultfd-based
mechanism impairs, by itself, the performances of the protected applications.
Note that, even if the GPU kernel is a component that can be easily replaced,
for instance by an implementation of another, more efficient cipher,
transferring the data of the protected pages between system RAM and GPU has an
intrinsic cost that could not be easily reduced in the current \Project\
implementation.

\begin{table}[!b]
    \begin{subtable}[b]{0.49\textwidth}
    \begin{center}
\begin{tabular}{|cc|r|r|}
    \cline{3-4}
    \multicolumn{2}{c|}{} & \multicolumn{2}{c|}{\textbf{aes}} \\
    \multicolumn{2}{c|}{} & \multicolumn{1}{c}{\textbf{\small Baseline}} & 
    \multicolumn{1}{c|}{\textbf{\small \Project}} \\
    \hline
\multirow{3}{*}{~\rotatebox{90}{\parbox{2.5em}{\textbf{Inst.}}}}
& \textbf{1} & 0.30 $\pm$ 0.03 & 35.23 $\pm$ 0.22 \\
& \textbf{2} & 0.30 $\pm$ 0.01 & 36.77 $\pm$ 0.16 \\
& \textbf{4} & 0.31 $\pm$ 0.01 & 39.91 $\pm$ 0.47 \\
\hline
\end{tabular}
\end{center}

    \end{subtable}
    \begin{subtable}[b]{0.49\textwidth}
    \begin{center}
\begin{tabular}{|cc|r|r|}
    \cline{3-4}
    \multicolumn{2}{c|}{} & \multicolumn{2}{c|}{\textbf{sha512}} \\
    \multicolumn{2}{c|}{} & \multicolumn{1}{c}{\textbf{\small Baseline}} & 
    \multicolumn{1}{c|}{\textbf{\small \Project}} \\
    \hline
\multirow{3}{*}{~\rotatebox{90}{\parbox{2.5em}{\textbf{Inst.}}}}
& \textbf{1} & 0.44 $\pm$ 0.01 & 4.58  $\pm$ 0.03 \\
& \textbf{2} & 0.46 $\pm$ 0.01 & 4.82  $\pm$ 0.03 \\
& \textbf{4} & 0.48 $\pm$ 0.01 & 5.28  $\pm$ 0.06 \\
\hline
\end{tabular}
\end{center}

    \end{subtable}
    \par\vskip1em
    \begin{subtable}[b]{0.49\textwidth}
    \begin{center}
\begin{tabular}{|cc|r|r|}
    \cline{3-4}
    \multicolumn{2}{c|}{} & \multicolumn{2}{c|}{\textbf{qsort}} \\
    \multicolumn{2}{c|}{} & \multicolumn{1}{c}{\textbf{\small Baseline}} & 
    \multicolumn{1}{c|}{\textbf{\small \Project}} \\
    \hline
\multirow{3}{*}{~\rotatebox{90}{\parbox{2.5em}{\textbf{Inst.}}}}
& \textbf{1} & 4.67 $\pm$ 0.03 & 64.35 $\pm$ 1.62 \\
& \textbf{2} & 4.74 $\pm$ 0.03 & 67.26 $\pm$ 1.23 \\
& \textbf{4} & 4.94 $\pm$ 0.01 & 73.37 $\pm$ 1.26 \\
\hline
\end{tabular}
\end{center}

    \end{subtable}
    \begin{subtable}[b]{0.49\textwidth}
    \begin{center}
\begin{tabular}{|cc|r|r|}
    \cline{3-4}
    \multicolumn{2}{c|}{} & \multicolumn{2}{c|}{\textbf{crypt}} \\
    \multicolumn{2}{c|}{} & \multicolumn{1}{c}{\textbf{\small Baseline}} & 
    \multicolumn{1}{c|}{\textbf{\small \Project}} \\
    \hline
\multirow{3}{*}{~\rotatebox{90}{\parbox{2.5em}{\textbf{Inst.}}}}
& \textbf{1} & 4.76 $\pm$ 0.02 & 12.52 $\pm$ 0.08 \\
& \textbf{2} & 4.81 $\pm$ 0.04 & 13.15 $\pm$ 0.14 \\
& \textbf{4} & 5.03 $\pm$ 0.02 & 15.04 $\pm$ 0.10 \\
\hline
\end{tabular}
\end{center}

    \end{subtable}
    \vskip1em
	\caption{%
    Wall-clock execution times of concurrent instances of the benchmarks,
    without \Project\ (``Baseline'') and with \Project\ using a sliding window
    of four pages.  Average times and standard deviations are in seconds.
    ``Inst.'' denotes the number of concurrent benchmark instances.  \Project\
    encrypts all private anonymous pages, including the stack.}
	\label{t:insts}
\end{table}

The \emph{aes} benchmark has a very large slowdown for sliding window size less
than 16. A typical AES implementation makes use of very large tables, which are
continuously and randomly accessed. The plot shows that the accesses fall both
into many \texttt{malloc()}ed pages and, to a lesser extent, into the stack.
Reducing the sliding window size from eight to four significantly increases the
slowdown, which means that the accesses to the encrypted pages are quite
random. GPU encryption roughly doubles the slowdown values.

The \emph{sha512} benchmark has negligible slowdown for sliding window sizes
greater than eight. The plot shows that the overhead is due mainly to anonymous
private pages not included in the stack, that is, stack handling does not
contribute a lot to the slowdown. In fact, the program uses several
\texttt{malloc()}ed pages as a buffer for data read from file, while it uses
the stack to store a small data structure for the digest inner state.
Decreasing the sliding window size from eight to four does not significantly
change the slowdown, which means that the replacement policy of the sliding
window is working fine: actually, the program tends to read sequentially the
pages in the buffer.  GPU encryption doubles the slowdown values.

The \emph{qsort} benchmark has been selected so to emphasize the effects of
stack encryption in some types of protected applications. The program sorts
2048 integers eventually using the C library function \texttt{qsort()}. This
function avoids recursion by storing on the stack some pointers to the array
partitions still to be sorted. Using a sliding window of four pages, page fault
handling causes a slowdown roughly seven when the stack is protected, and the
GPU encryption doubles this value. On the other hand, if stack is not included
in the protected pages, the slowdown is always negligible.

Similarly, the \emph{crypt} benchmark has a significant slowdown only when the
stack is encrypted. In fact, the \texttt{crypt\_r()} function is invoked on
data placed on the stack. The slowdown caused by GPU encryption, by itself, is
roughly equal to the slowdown caused by page fault handling. The overhead of
\Project\ is quite small, for every sliding window size.

\begin{figure}[t]
	\centering
	\begin{subfigure}[b]{0.49\textwidth}
		\includegraphics[width=\textwidth]{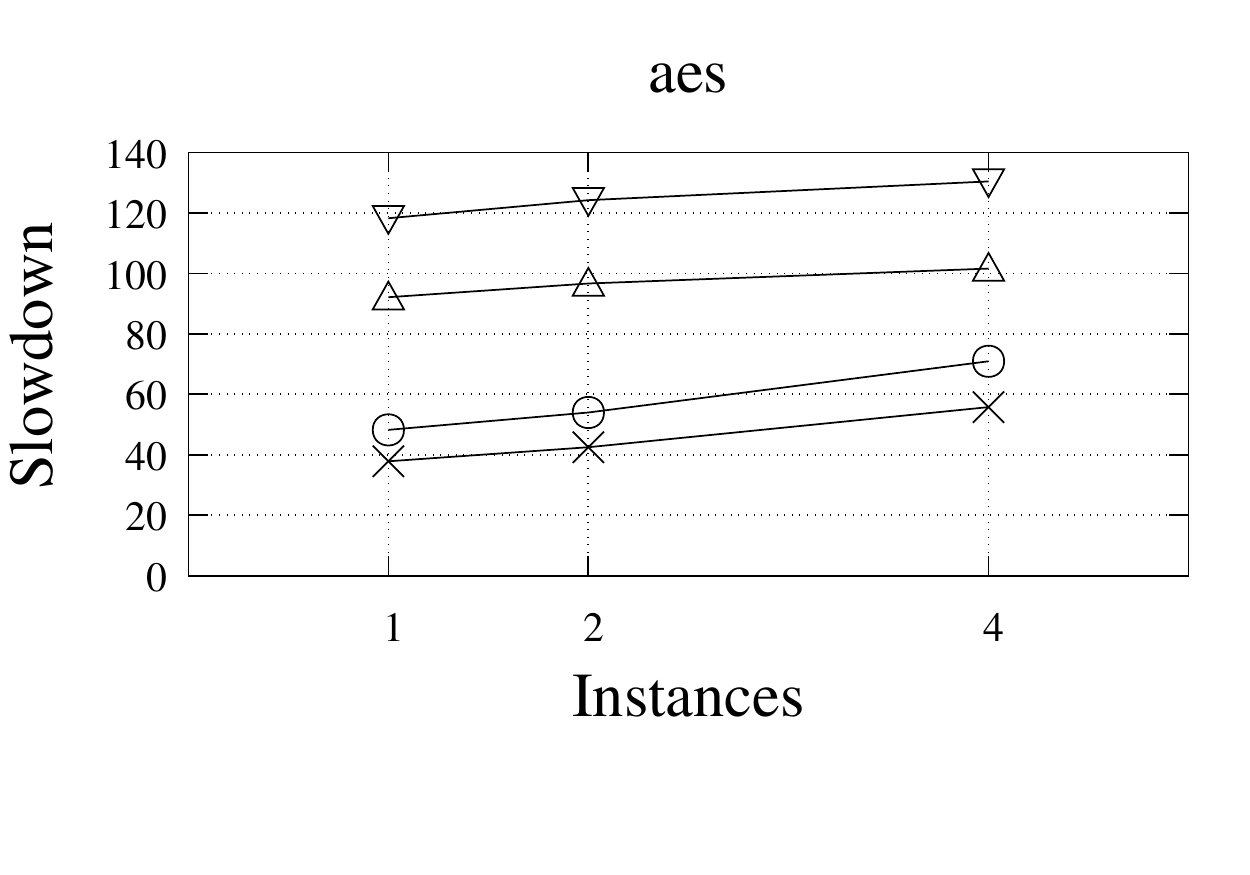}
	\end{subfigure}
	\begin{subfigure}[b]{0.49\textwidth}
		\includegraphics[width=\textwidth]{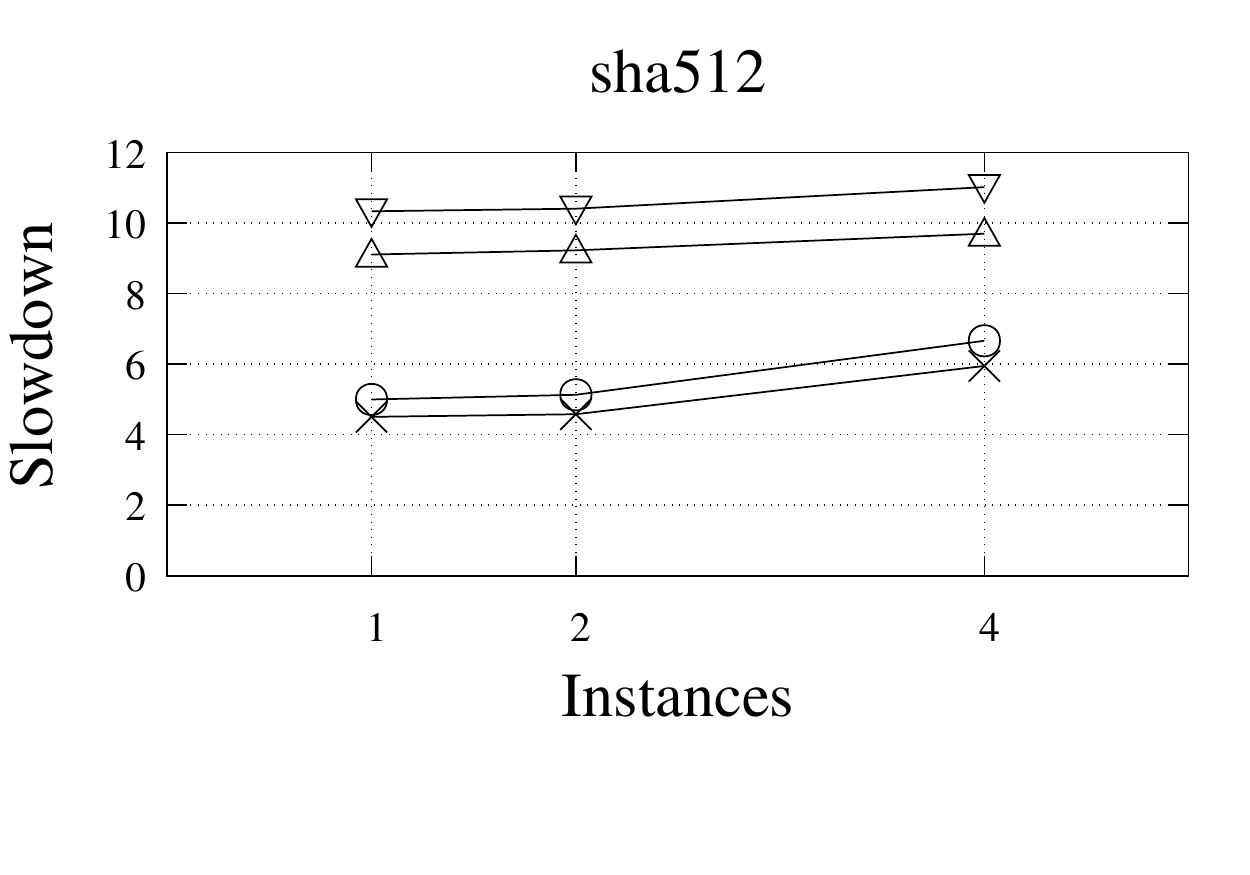}
	\end{subfigure}
    \vskip-2em\includegraphics[height=1.9em]{legend.pdf}\vskip-.5em
	\begin{subfigure}[b]{0.49\textwidth}
		\includegraphics[width=\textwidth]{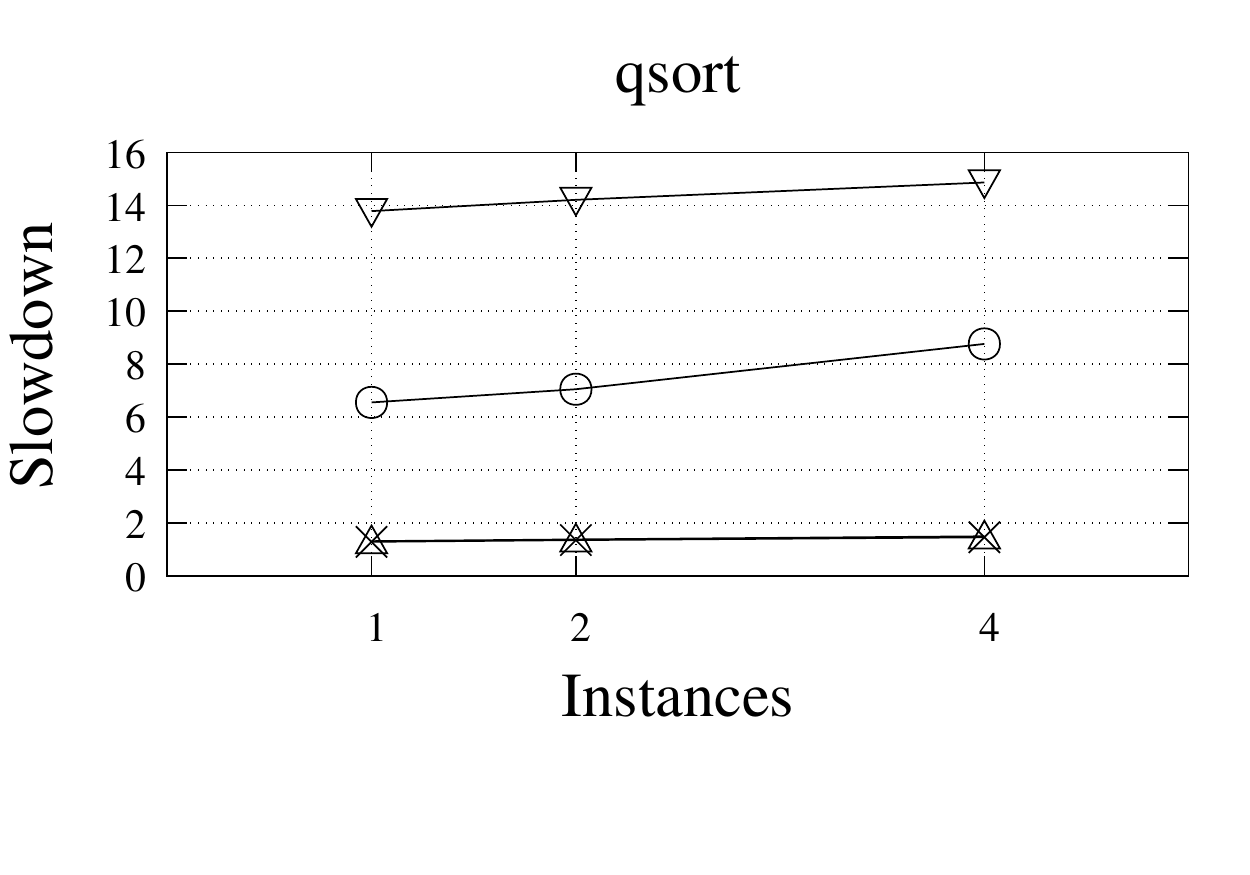}
	\end{subfigure}
	\begin{subfigure}[b]{0.49\textwidth}
		\includegraphics[width=\textwidth]{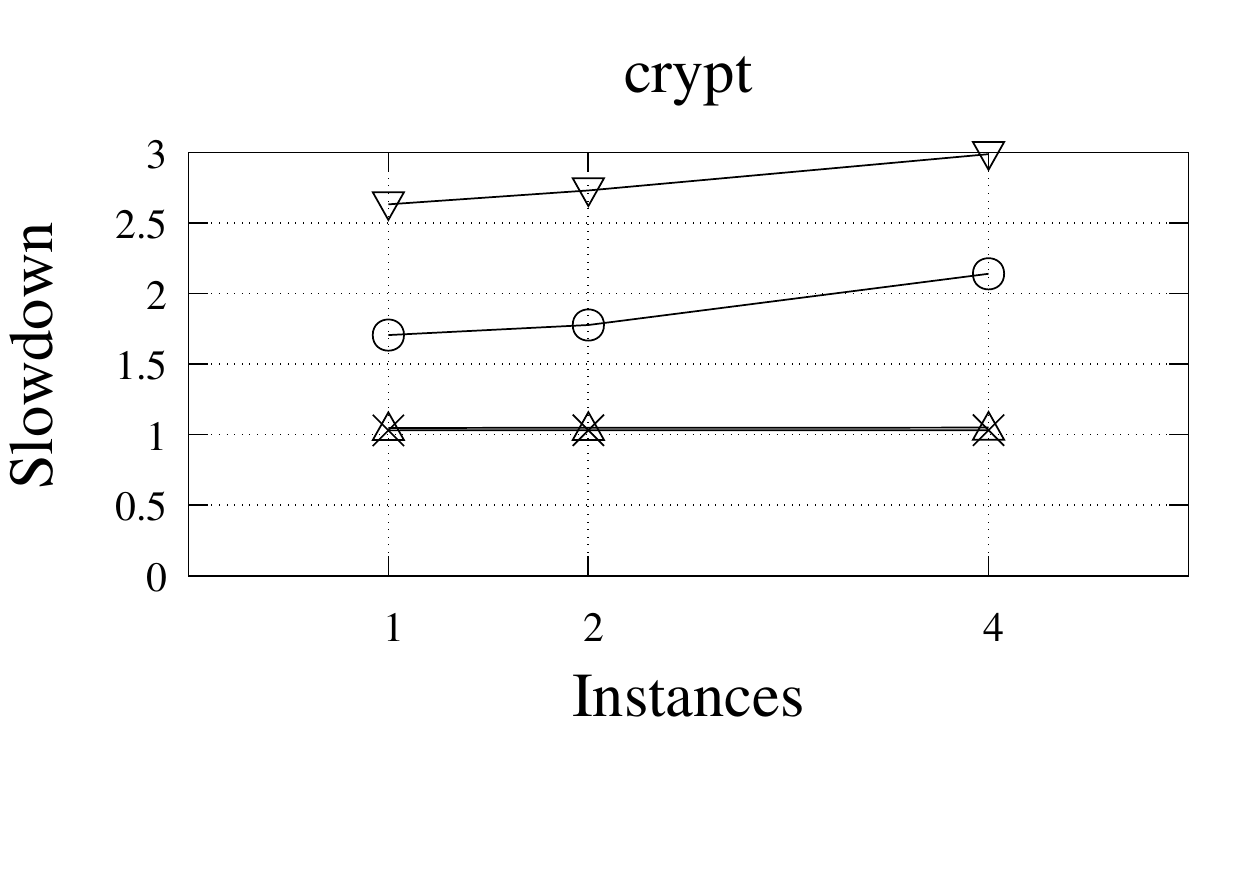}
	\end{subfigure}
    \vskip-2em
    \caption{Slowdowns relative to baseline (no \Project) varying the number of
    benchmark instances. \Project\ protects all private anonymous pages
    including the stack (``Stack'') or excluding it (``no Stack''), with a
    sliding window of 4 pages.  Encryption is performed by the GPU
    (``GPU'') or not done at all (``no GPU'').}
    \label{f:insts}
\end{figure}

We also run another set of benchmarks to verify how \Project\ scales up when
the number of protected clients increases. Each benchmark launches one, two, or
four instances of the program at the same time; observe that our test machine
has only four physical cores.  As shown in Table~\ref{t:insts}, average
execution times have been collected with and without \Project\ protection, with
sliding window size equal to four pages and stack encryption enabled.
Figure~\ref{f:insts} shows the slowdowns of the average execution times with
respect to the baseline, that is, execution without \Project\ protection.
According to the plots, there is a limited increase of the slowdown when the
number of concurrent instances grows, both for page fault handling and for GPU
encryption.

\section{Conclusions}
\label{s:conc}

Memory encryption is an effective solution against memory disclosure attacks,
in which an attacker could access and dump system RAM contents, in order to
extract any sort of sensitive data.
%
In this article we presented \Project{}, which is a novel approach to
software memory encryption for user-mode applications that uses a GPU as a
secure key store and crypto processor. By ensuring that the key and the
internal state of the cipher are stored into GPU hardware registers, \Project{}
guarantees that this sensitive information are never leaked to system RAM;
therefore, the attacker cannot get meaningful data from a system dump unless
the encryption cipher is broken.

Compared to all other proposals for memory encryption frameworks,\linebreak
\Project\ is implemented by a user-mode daemon and does not require patches to
the operating system kernel. Moreover, user applications do not require source
code changes, recompilation, or code instrumentation, hence \Project\ can
protect even applications for which only the executable code is available.

Functional tests and security analysis suggest that \Project\ is an effective
mitigation of Cold Boot attacks aimed at system RAM. Performance measures on a
prototype implementation show how the \Project\ overhead heavily depends on the
chosen configuration and clients' memory access patterns. Moreover, the current
implementation can be significantly improved, for instance by implementing
selective page encryption, by optimizing the GPU kernel implementation, or by
introducing some mechanisms that start encrypting pages in the sliding window
``in background'' as the number of free slots becomes lower than a predefined
threshold.

\section*{Acknowledgments} We gratefully thank Emiliano Betti
for his valuable suggestions, support, and encouragements.
The material presented in this paper is based upon work partially supported
by Epigenesys s.r.l.. Any opinions, findings, and conclusions or
recommendations expressed in this publication are those of the
authors and do not necessarily reflect the view of Epigenesys s.r.l..

%
%
\bibliographystyle{splncs04}
\bibliography{memshield.bib}
\end{document}